# Organic room-temperature polariton condensate in a higher-order topological lattice

*Christoph Bennenhei[1], Hangyong Shan[1], Marti Struve[1], Nils Kunte[1], Falk Eilenberger[2], Jürgen Ohmer[3], Utz Fischer[3], Stefan Schumacher[4], Xuekai Ma[4], Christian Schneider[1], Martin Esmann[1,*]*

[1]*Institute of Physics, School of Mathematics and Science, Carl von Ossietzky Universität Oldenburg, 26129 Oldenburg, Germany*

[2]*Institute of Applied Physics, Abbe Center of Photonics, Friedrich Schiller University Jena, 07743 Jena, Germany; Fraunhofer-Institute for Applied Optics and Precision Engineering IOF, 07743 Jena, Germany; Max-Planck-School of Photonics, 07743 Jena, Germany*

[3]*Department of Biochemistry, University of Würzburg, 97074 Würzburg, Germany*

[4]*Department of Physics, Center for Optoelectronics and Photonics Paderborn (CeOPP), and Institute for Photonic Quantum Systems (PhoQS), Paderborn University, 33098 Paderborn, Germany*

**ABSTRACT**

Organic molecule exciton-polaritons in photonic lattices are a versatile platform to emulate unconventional phases of matter at ambient conditions, including protected interface modes in topological insulators. Here, we investigate bosonic condensation in the most prototypical higher-order topological lattice: a 2D-version of the Su-Schrieffer-Heeger (SSH) model, supporting both 0D and 1D topological modes. We study fluorescent protein-filled, structured microcavities defining a staggered photonic trapping potential and observe the resulting first- and higher-order topologically protected modes via spatially resolved photoluminescence spectroscopy. We account for the spatial mode patterns by tight-binding calculations and theoretically characterize the topological invariants of the lattice. Under strong optical pumping, we observe bosonic condensation into the topological modes. Via interferometric measurements, we map the spatial first-order coherence in the protected 1D modes extending over 10μm. Our findings pave the way towards organic on-chip polaritonics using higher-order topology as a tool for the generation of robustly confined polaritonic lasing states.

**INTRODUCTION**

Organic molecules hosting tightly bound Frenkel excitons are a particularly promising and easily accessible platform for the realization of cavity exciton-polariton devices [1–4]. The exciton Bohr radius of roughly the size of a single molecule yields binding energies on the order of 1 eV, making these excitons stable at ambient conditions. Polariton lasing [5,6] in organic matter was first demonstrated in polymers [3] followed by other molecules [2]. Fluorescent proteins are a more recent addition to this group of low-cost, flexible, active materials for room-temperature polariton lasing and Bose-Einstein condensation [7–11]. In particular they feature excellent quantum efficiency and, due to their structure consisting of an exterior β-barrel around an interior chromophore, their emission wavelength is widely tunable with virtually identical chemical properties.

Engineering of the spatial profile of polariton condensates is most commonly realized by structuring the photonic cavity, e.g. by plano-concave photonic potentials that confine the optical field in all three dimensions of space [8]. Based on these advances, the formation of room-temperature polariton condensates in coupled 1D arrays of spherical cap-shaped, plano-concave optical cavities in topologically trivial and non-trivial arrangements has been demonstrated [10].

Employing the non-trivial topology of the photonic band structure to control the confinement of polaritonic modes at the interface between topologically different domains is particularly appealing for the generation of robust polariton lasers at room-temperature on chip. While lasing from 0D topological defects in a 1D bulk (i.e. a first order topological insulator) has been observed in various organic and inorganic polariton systems [10,12,13] only very recently a first experimental demonstration of second-order topological polariton lasing (i.e. lasing from a 0D corner mode embedded in a 2D bulk) was discussed in a perovskite lattice [14]. Here, we employ organic polaritons with excitons in the fluorescent protein mCherry [15] coupled to a 2D lattice of plano-concave microcavities hosting a 0D topological defect as well as 1D line-defects. We theoretically characterize the non-trivial bulk topology and show in spatially resolved photoluminescence (PL) measurements that our lattice supports both 1D and 0D defect modes. Under strong optical pumping, we observe a pronounced optical non-linearity in the input-output response indicating the onset of polariton lasing from the topologically protected states. We furthermore observe clear indications of the formation of topologically protected 1D polariton condensates through mapping of the $g^{(1)}$ coherence along the topologically protected 1D line-defects. These results further advance the field of on-chip polaritonics towards topologically robust control over polariton lasers with multiple dimensionalities.

**RESULTS**

Figure 1 illustrates the topologically different phases of the Su-Schrieffer-Heeger model in one (panel (a)) and two dimensions (panel (e)). For a dimerized lattice composed of centro-symmetric unit cells in a single band, one particle, tight-binding approximation, a topologically non-trivial phase arises if the intra-cell nearest-neighbor hopping integral $w$ is smaller than the inter-cell hopping $v$. In the 1D case, this non-trivial topology can for example be expressed via the winding number of the chiral momentum-space Hamiltonian in the basis of the spin-1/2 Pauli matrices [16,17]. By virtue of the bulk-boundary correspondence, a finite 1D chain in the non-trivial phase exhibits localized states at the monomeric termination points interfacing the topologically trivial vacuum. These states decay exponentially into the bulk and exhibit complete sublattice polarization, i.e. the wave function of the topological interface mode has non-zero components on only one sublattice of the SSH chain. Due to the chirality of the Hamiltonian, the eigenenergy spectrum is symmetric and thus the defect state remains pinned at the bandgap center, protected from any fluctuations that do not break chirality.

The (non-trivial) topology of a 2D lattice can be characterized by its bulk polarization. We calculate this quantity via the well-established Wilson-loop approach [14] (Supplementary Information Section 1). In the present case, the polarization acquires $P = 1/2$ (P can only be 1/2 or 0) for a dimension along which the lattice exhibits non-trivial topology while it is $P = 0$ otherwise. This results in four possible unit-cells in the 2D case as shown in Fig. 1 (e), with non-trivial topology along none, one, or both directions of space.

Figure 1(b,f) illustrates our specific implementations of the lattice geometries in one and two dimensions. In the 1D chain, we interface a topologically non-trivial phase on the left with a topologically trivial phase on the right. The dashed line marks the domain boundary while the red site labels the resulting topological monomeric defect, which is one dimension lower that the embedding bulk. In 2D, we interface the four possible topologically different unit cells as sketched, resulting in two 1D line-defects (red, one dimension lower than the bulk) forming a 0D monomeric defect at their intersection (green, two dimensions lower than the bulk).

We perform effective tight-binding simulations of the two structures resulting in the eigenenergy spectra plotted in Fig. 1(c,g). Parameters were $2meV$ for the strong and $0.9 meV$ for the weak hopping links. For the 1D case, we retrieve the well-established cosine-shaped band with a topological band gap around zero energy due to the staggered hopping and the gapped topological mode at zero-energy. For the 2D case, it has been shown that if only isotropic nearest-neighbor hopping is included [14,18] the spectrum remains symmetric in energy but no global band gap around zero is present such that any state with a dominant amplitude contribution on the 0D defect can couple to energetically (near-)degenerate bulk modes. This degeneracy may be lifted through at least two effects: 1. The inclusion of a next-nearest neighbor (NNN) hopping amplitude $a$ [14] (cf. Supplementary Section 1) that represents additional diagonal hopping links between non-trivial unit cells. 2. A small anisotropy in the hopping links, i.e. $v$ and $w$ being slightly different along $x$ and $y$ [18,19]. For the theoretical calculation in Fig. 1(g) we included a NNN hopping of $a = 2meV$ [14] resulting in a globally gapped 0D state slightly above zero energy (violet symbol).

In Figure 1(d,h), we show scanning electron microscope (SEM) images of the 1D and 2D photonic lattices implemented following the concept in Fig. 1(b,f) by producing arrays of coupled plano-concave organic microcavities. The image shows overlapping indentations in the shape of spherical caps with diameter $d = 3 - 5 \mu m$ and 77-155nm depth, which we fabricate into a glass substrate via ion beam lithography. Hopping links correspond to a center-to-center distance between neighboring indentations of $0.65d$ for the weak and $0.52d$ for strong links. The indentations serve as a template for depositing a SiO$_2$/TiO$_2$ distributed Bragg reflector (DBR) with 8 pairs and a center of the optical stop band at 610nm (see methods section for details). A solution of the fluorescent molecule mCherry is then laminated between this structured DBR and a second, planar DBR to form a lattice of resonant, plano-concave microcavities. (see Methods section and Supplementary Section S4 for details).

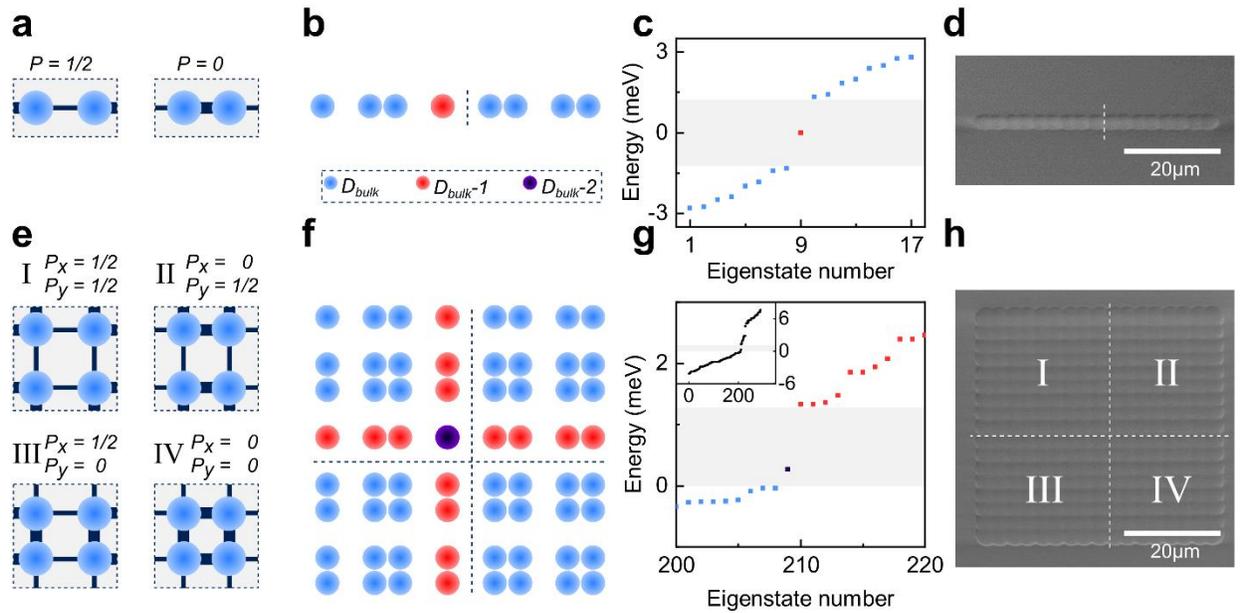

**Figure 1**. **a,e** Unit cells of 1D (a) and 2D (e) implementations of the Su-Schrieffer-Heeger model. A topologically non-trivial phase supporting protected edge modes arises if the intercell hopping is larger than the intracell hopping. Thick (thin) lines indicate strong (weak) hopping. **b,f** Sketch of the implementations as used in the experiments for a 1D SSH chain (b) with a 0D defect (red) and of a 2D SSH lattice (f) with line defects (red) and a monomeric defect (violet). Dashed lines mark the interfaces between domains of different bulk polarization $P_{x/y}$ calculated via the Wilson loop approach. **c,g** Calculated tight binding eigenenergies of the 1D (c) and 2D lattice (g) for $2meV$ strong and $0.9meV$ weak hopping. The 1D case shows a gapped defect state. The 2D case (bottom) shows dispersive gapped 1D states and a gapped 0D state since an additional next-nearest neighbor hopping of $a = 2meV$ is included. The color of the plotted eigenstates represents the dimensionality with respect to the bulk. In (c) red corresponds to the monomeric defect, while in (g) red represents the 1D states and violet the 0D state. **d,h** Scanning electron microscope (SEM) images of the 1D and 2D photonic lattice implemented by producing indentations in the shape of spherical caps with $d = 4\mu m$ in one distributed Bragg reflector (DBR) of the organic microcavity via ion beam lithography.

As a first step, we analyze the optical behavior of a single line segment of the 2D lattice, which is a one dimensional SSH chain featuring a central domain boundary as shown in Fig. 1 (top row). Figure 2 shows experimental PL measurements on a 1D SSH chain of mCherry filled plano-concave microresonators. Under non-resonant cw-excitation at 532nm localized on the central topological defect, we record the energy-resolved real-space PL map in panel (a, bottom). Two Bloch bands separated by a topological band gap around 1.88eV arise from the (anti-)symmetric hybridization of s-type Laguerre-Gaussian orbitals localized in each microresonator. Based on a coupled oscillator model fitted to the measured polariton dispersion of a planar microcavity [8] (see Supplementary Fig. S1) we deduce a vacuum Rabi splitting of 210 meV and a red-detuning of 220 meV from the exciton at 2.1 eV for the band structure in Fig. 2. Within the topological band gap resides the topologically protected interface mode, for which we show a cross-section in the upper part of the panel. We measure an exponential decay length into the bulk of ~5µm on the intensity level and a sub-lattice polarization of $S = \frac{I_A - I_B}{I_A + I_B} \approx 0.72$ obtained by measuring PL intensities $I_A$ and $I_B$ on the A and B sub-lattices for the two unit cells bounding the topological interface. By changing from real- to momentum-space, we obtain the PL dispersion relation in Fig. 2(b). The topological interface mode is localized at the edges of the first Brillouin zone with a band of (anti-)binding s-modes below (above). Bands from higher order p-modes both along and across the SSH chain are also visible.

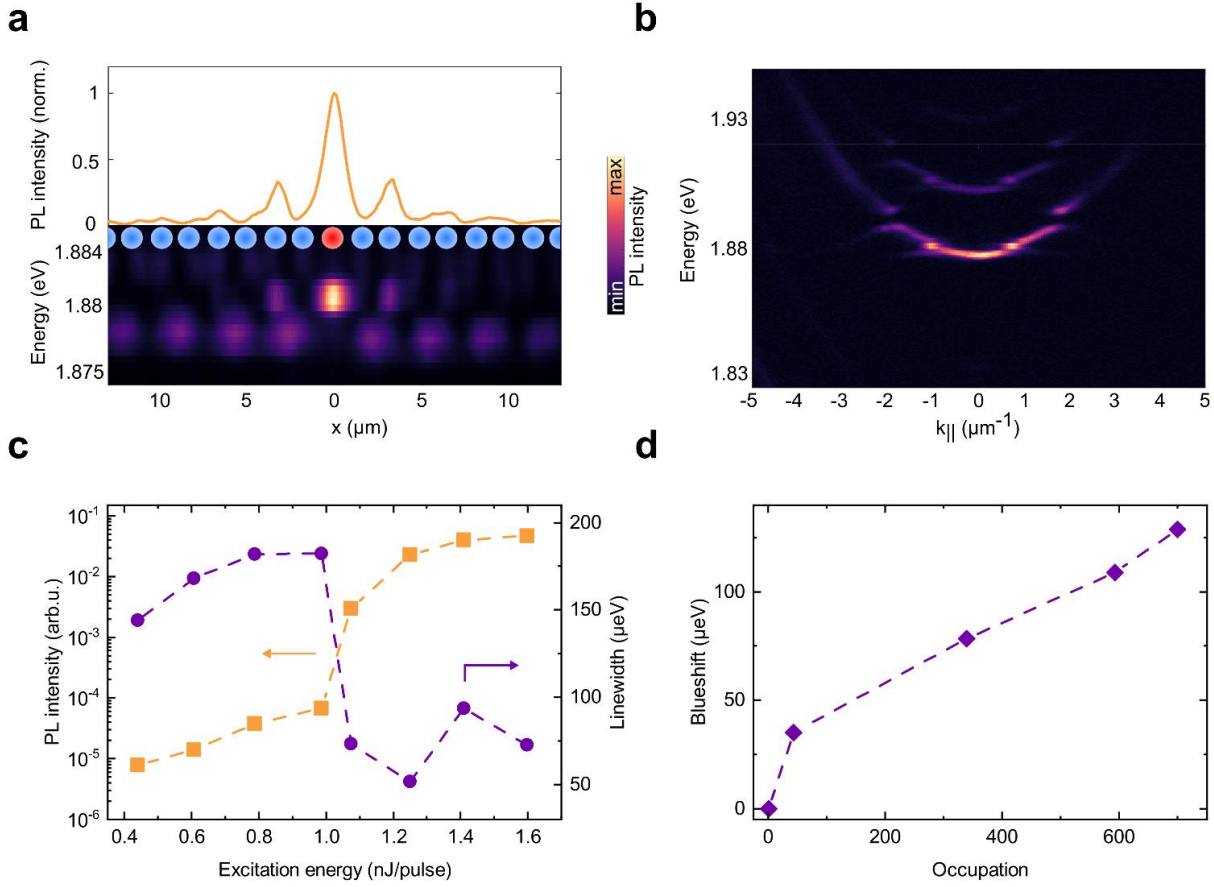

**Figure 2**. **a** Bottom: Energy-resolved real-space photoluminescence of a 1D SSH chain under non-resonant cw excitation at 532nm. The spectrally gapped, topologically protected 0D mode is exponentially localized on the central topological defect, with a measured sub-lattice polarization of $S \approx 0.72$. Top: Spatial cross-section extracted from the bottom panel around 1.88eV. **b** Momentum- and energy-resolved photoluminescence of the 1D SSH chain. The dispersive s-band features a topological gap with the defect mode localized in momentum space at the edges of the first Brillouin zone ($k_\parallel \approx \pm 1 \mu m^{-1}$). Further bands arise from p-orbitals along x and y. **c** Input-output characteristic (orange) of the 0D edge defect under pulsed non-resonant excitation (532nm, 7ns pulse duration). A clear non-linearity is observed at pulse energies around 1nJ, accompanied by a step-like collapse of the emission linewidth (violet) to the instrument resolution limit around 70μeV. At even higher pulse energies, the emission shows typical saturation behavior. **d** Blueshift of the photoluminescence above lasing threshold. Beyond an average polariton occupation of $\langle n \rangle = 2$, we find a shift of $\sim 4\mu eV$ per polariton.

Under quasi-steady state excitation conditions (532nm, 7ns pulse duration, centered on the central defect) we record the input-output characteristic of the topological interface state shown in panel (c). For pulse energies above $P_{th} \approx 1nJ$, we observe a clear non-linearity, accompanied by a collapse of the polariton line width down to the spectral resolution limit of the optical detection system. After $3P_{th}$ the PL intensity saturates. Due to phase space filling, we also observe a blueshift of the polariton emission with pump power (Fig. 2(d)) that amounts to ~4µeV per polariton in average occupation. Here, we set the average occupation at the threshold to unity and use that average occupation above threshold is proportional to the detected PL intensity. We interpret the sum of these observations as clear signatures of a transition to polariton lasing in the high polariton density regime.

This SSH chain constitutes a 1D section of the 2D lattice, the study of which is detailed in the following.

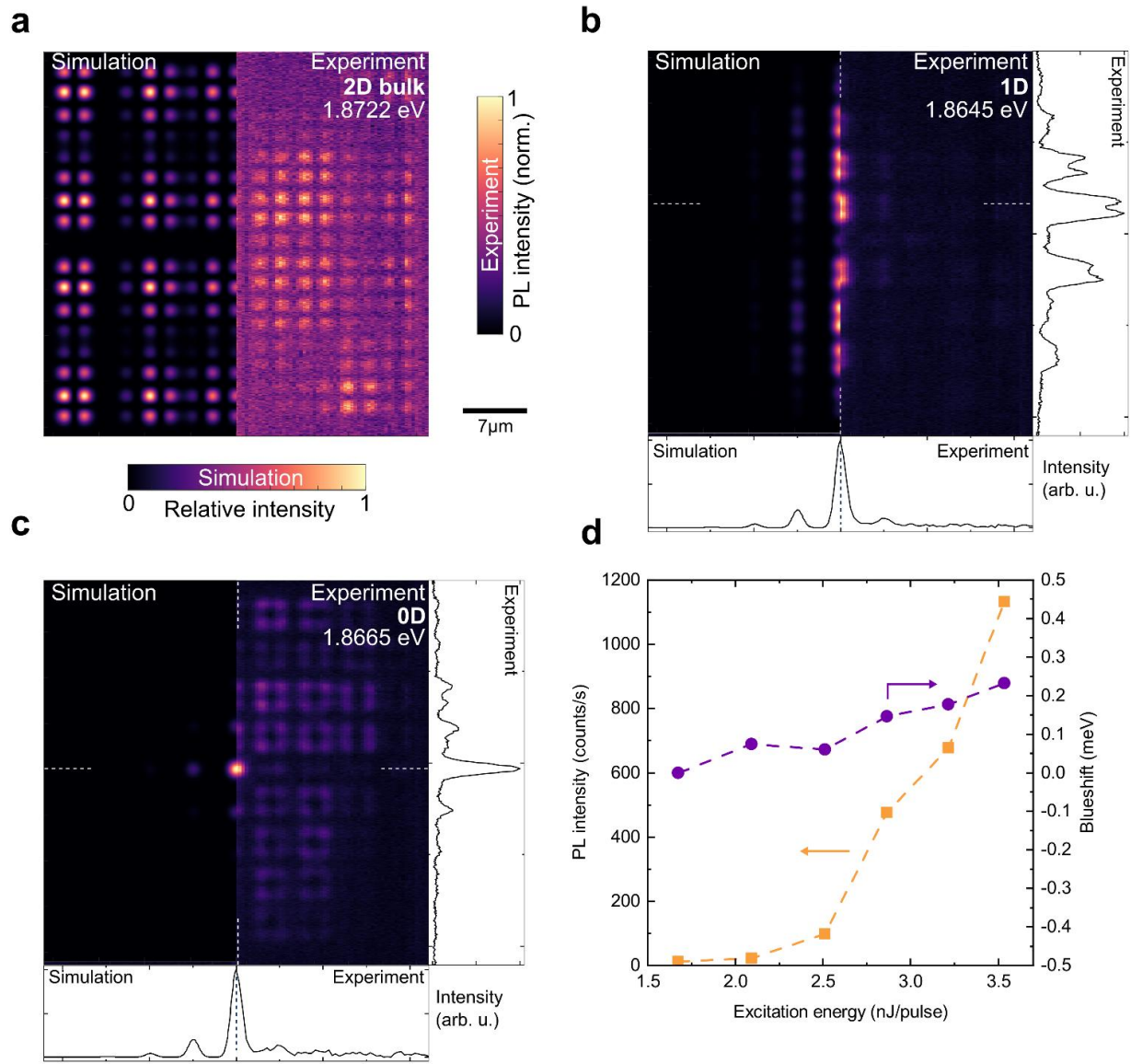

**Figure 3**. **a-c** Spatial modes of a polaritonic 2D SSH lattice . The left half of each panel shows the spatial intensity distribution of an eigenmode calculated with a tight-binding model including a 3% anisotropy in the hopping links (see Supplementary Section S3). The right half of each panel shows corresponding experimental photoluminescence images of these spatial modes obtained from hyperspectral tomography (see Methods section) taken under local off-resonant cw excitation at 532 nm on the central monomeric defect. Panel **a** shows an exemplary 2D bulk mode (antibinding), **b** shows a topologically protected 1D channel mode and **c** the 0D mode at the central monomeric defect of the lattice. **d** Input-output characteristic (orange) and accompanying blueshift (violet) of the polariton lasing transition of a 2D SSH lattice under off-resonant excitation at 532 nm with 7 ns laser pulse duration localized on the central defect.

Our optical study on the 2D SSH lattice is carried out via energy-resolved real-space PL mapping of the lattice depicted in Fig. 1(h) via non-resonant cw excitation on the central defect. This mapping is achieved by a tomographic reconstruction of a 2D map from 1D sections analogous to the mapping in Fig. 2. In Fig. 3 (a-c) we present three constant-energy sections of this dataset in the right halves of the panels and compare them to real-space eigenstates from a tight-binding calculation, convoluted with an s-type orbital to achieve the simulated intensity distributions. Panel (a) shows a representative bulk state from an anti-binding band. Panel (b) shows a topologically protected dispersive, 1D interface state localized on the vertical line defect of the lattice. In x-direction, this mode exhibits the expected sub-lattice polarization (see cross-section). Panel (c) shows the higher-order topological 0D monomeric state localized at the central intersection point of the horizontal and vertical line defects. As visible in the cross-sections, this state exhibits the expected sub-lattice polarization both along x and y. To accurately reproduce these experimental findings in the tight-binding simulations, we introduce a 3% anisotropy in the staggered hoppings along x and y [18], i.e. $w_x/w_y = v_x/v_y = 1.03$ (Supplementary Section S3). Such an anisotropy most likely arises from the spatial gradient in cavity thickness and leads to a 0D state localized in a global energy gap of 80µeV width. Conversely, our simulations clearly show that in the absence of both an anisotropy and dominant NNN hopping any monomeric 0D state hybridizes with close-by bulk states and most importantly does not exhibit the expected sub-lattice polarization of the SSH state [14]. Our experimental observations of both localization and sublattice polarization thus evidence that the higher-order 0D mode is energetically gapped.

Under non-resonant quasi-steady state excitation (7 ns pulse duration, 532 nm) centered on the central defect, we detect a clear non-linearity in the input-output characteristic presented in Fig.

3(d) with a threshold of $P_{th} \approx 2.5\, nJ$. The resulting blueshift shown as the inset amounts to $0.2\, meV$.

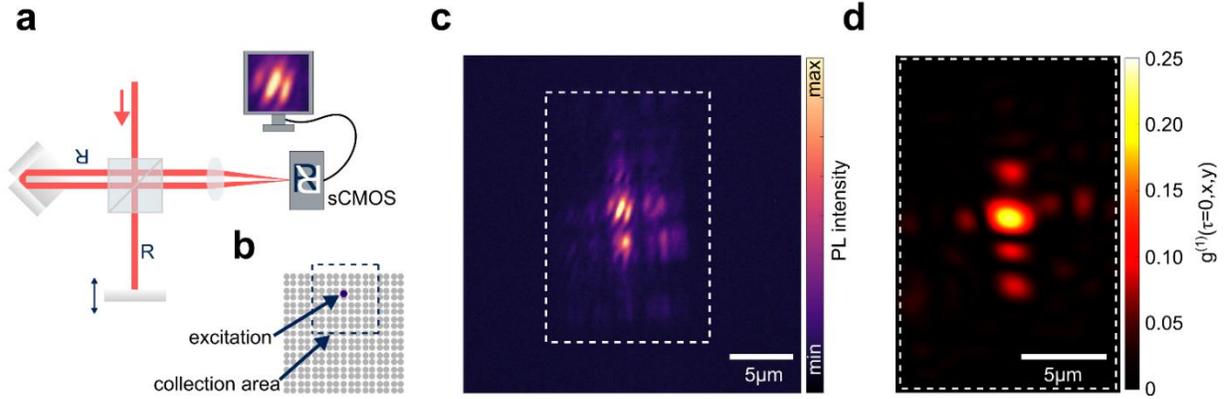

**Figure 4**. **a** Schematic of the Michelson interferometer with a retroreflector in the reference arm, flipping the image along two spatial axes and a flat mirror in the delay arm on a translation stage. **b** Position of the excitation (violet dot) and collection area (dashed square) within the 2D SSH lattice. **c** Corresponding interference image at zero time delay between the interferometer arms. **d** False color representation of the first order correlation function $g^{(1)}(\tau = 0, x, y)$ extracted from the interference image shown in (c).

Having established room-temperature polariton lasing in the 2D SSH lattice, we now show that the 1D channel modes even support long range order in the presence of topological protection. To this end, we locally pumped a polariton condensate into one of the 1D topological channel modes and interferometrically measured its spatial first order correlation function $g^{(1)}(\tau = 0, x, y)$ along the channel. This is achieved by the setup sketched in Fig. 4(a) composed of a Michelson interferometer with a retroreflector in one arm tuned close to zero temporal delay. Panel (b) shows the location of the pulsed off-resonant pump (violet) and the area from which we collect the monochromatic emission of the condensate (dashed violet). Panel (c) shows an image of the polariton condensate emission superimposed with its spatially inverted copy at the interferometer exit measured at $3.5\, nJ$ excitation pulse energy. A slight overall offset in wave vector direction between the two copies results in pronounced interference fringes. From this pattern and the emission images of two arms, we extract [5,20–22] a spatial coherence map inside the dashed rectangle in (c) and plot it as panel (d). Along the 1D topological edge channel, the coherence extends over a distance of ~10μm, i.e. well beyond the size of the localized pump spot, which confirms the macroscopic phase coherence in polaritons emission.

**CONCLUSION**

We have experimentally investigated bosonic condensation in the most prototypical higher-order topological lattice, a 2D-version of the Su-Schrieffer-Heeger (SSH) model. This lattice supports both 0D and 1D topological defect modes. We have experimentally studied a realization of this lattice in coupled mCherry-filled DBR microcavities and observed the resulting topological defect modes of different dimensionality. A comparison to tight-binding simulations revealed that the observed near-complete sub-lattice polarization of the 0D state is a clear indication of an energetically gapped higher-order topological interface mode. In the high-density regime of exciton-polaritons, we observed bosonic condensation in the topologically protected interface modes. In particular, interferometric measurements revealed long-range order in the presence of topological protection via first-order coherence extending over a length of ~10μm in topological 1D channels. These findings are an important step towards organic on-chip polaritonics at ambient conditions using higher-order topology as a tool for the generation of robustly confined lasing states and bosonic condensates.

**METHODS**

The staggered photonic trapping potentials investigated in this work are realized by a structured optical microcavity filled with fluorescent proteins. Figure S3 shows a schematic of the cavity fabrication. First, indentations in the shape of spherical caps, with a depth of $76 - 155\ nm$ and diameters of $d = 3 - 5\ \mu m$, are milled into a glass substrate using a focused ion beam ($Ga^+$, FEI Helios NanoLab 600i Dual Beam Microscope). The photonic chains and lattices are formed by overlapping the indentations, where the overlap is chosen as a center-to-center distance of $0.52d$ for the strong hopping links and $0.65d$ for the weak hopping links. Next, dielectric mirrors are fabricated by evaporating alternating quarter-wave-layers of $SiO_2$ and $TiO_2$ on the structured substrate (8 pairs) and a planar substrate (10 pairs), with layer thicknesses chosen such that the resulting DBR stopband is centered around 610nm. The mirrors are laminated to form an optical microcavity with the cavity volume completely filled by the fluorescent protein mCherry. The protein is prepared in solution of 175 g/L following the recipe described in [23] for the protein tdTomato, followed by an additional step of adding 5% vol glycerol at the end of the cleaning process. The cavity is assembled by drop casting 5 µl of the mCherry solution on the patterned DBR and left to dry for 180 s to enhance the lamination properties. The planar mirror is then placed onto the mCherry layer to close the cavity. A weight of 50 g is placed on the cavity and it is left to dry for 48 h at room-temperature in a temperature stabilized environment. This results in cavity lengths ranging from ~500 nm to ~4500 nm.

The experimental setup used for the spectral characterization of the momentum-space of the topological chains is the same Fourier imaging setup as described in [24]. The structures are excited by an off-resonant continuous wave DPSS laser emitting at 532 nm coupled into the beam path through the reflection pathway of a beam splitter (BS013, Thorlabs) and focused by a microscope objective with an extra-long working distance (Mitutoyo Plan Apo NIR HR 0.65NA) to a spot size of ~3 μm. The photoluminescence emission is collected in reflection geometry by the microscope objective (f1=4mm) through the beam splitter. The back-focal plane of the microscope objective is imaged onto the entrance slit of a spectrometer (Andor SR-500i-A-SIL with an Andor iKon-M 934 CCD camera) using a confocal set of lenses (f2=300 mm, f3=200 mm, f4=300 mm f5=400 mm), resulting in emission angle-resolved spectra. The excitation laser is removed in front of the spectrometer by a 550 nm long-pass filter.

For the spectral characterization of the real space emission of the topological chains and lattices, lens f4 is taken out of the beam path, changing the Fourier imaging system into a real space imaging system. The energy resolved spatial emission patterns are recorded by a tomography, for which the last lens in front of the spectrometer (f5) is moved laterally in equidistant steps, thereby imaging different parts of the structure onto the spectrometer slit. The images are then stitched from the recorded spectra. For the polariton lasing experiment a Q-switched laser (CNI Laser MPL-III-532-20nJ) with 7 ns pulse length emitting at 532 nm instead of the continuous wave laser. For the spatial coherence measurement, a Michelson interferometer (as shown in Figure 4(a), with retroreflector PS976M-B and beam splitter BSW10R, both Thorlabs) is inserted in the beam path between lenses f3 and f4. After checking for monochromaticity with the spectrometer, the overlapping images from the interferometer are imaged on a sCMOS camera (Zyla 5.5, Andor) to enhance the collection efficiency.

**AUTHOR INFORMATION**

**Corresponding Author**

*m.esmann@uni-oldenburg.de

**Author Contributions**

The manuscript was written through contributions of all authors. All authors have given approval to the final version of the manuscript.

**ACKNOWLEDGMENT**

The authors acknowledge support by the German research foundation (DFG) via the project SCHN1376 13.1. Financial support by the Niedersächsisches Ministerium für Wissenschaft und Kultur ("DyNano") is gratefully acknowledged. This work was funded by the Federal Ministry of Education and Research (BMBF) under grant numbers 13XP5053A (NanoscopeFutur-2D) and the European Social Funds via the Federal State of Thuringia (Grant ID 2021FGI0043). M.E. acknowledges support by the University of Oldenburg through a Carl von Ossietzky Young Researchers' Fellowship. S.S. and X.M. acknowledge support from the DFG via grant No. 519608013. The authors thank Dr. Heiko Knopf for the fabrication of the DBR mirrors.

# SUPPORTING INFORMATION

## Supplementary section S1: Topological phases of the 2D SSH model

In this section, we theoretically characterize the topologically non-trivial properties of our 2D lattice shown in Fig. 1 (bottom) of the main text. To this end, we decompose the lattice into four domains as shown in Fig. 1 of the main text, and calculate the bulk polarization of the four different unit-cells in these areas. Besides the nearest-neighbor (NN) coupling, the next-nearest-neighbor (NNN) coupling is also included in our work, since the NNN coupling results in a topologically gapped 0D state as previously reported in [1]. The unit-cells can thus be described by the following tight-binding model in k-space

$$H = H_{NN} + H_{NNN}$$
$$= \begin{pmatrix} 0 & H_{12} & H_{13} & H_{14} \\ H_{12}^* & 0 & H_{23} & H_{24} \\ H_{13}^* & H_{23}^* & 0 & H_{34} \\ H_{14}^* & H_{24}^* & H_{34}^* & 0 \end{pmatrix} \quad (1)$$

with NN coupling elements

$$H_{12} = H_{34} = J_x + J'_x e^{ik_x a}$$
$$H_{13} = H_{24} = J_y + J'_y e^{-ik_y a} \quad (2)$$

Here, $J_{x,y}$ ($J'_{x,y}$) represents the intra-cell (inter-cell) coupling strength along $x, y$ directions, respectively. $a$ is the lattice constant of the unit-cells. Note that in the main text, these notations correspond to $J_{x,y} = w_{x,y}$ and $J'_{x,y} = v_{x,y}$. The elements that denote the NNN coupling, $H_{14}$ and $H_{23}$, are different for the different unit-cells as listed below:

|  | unit-cell I | unit-cell II | unit-cell III | unit-cell IV |
|---|---|---|---|---|
| $H_{14} =$ <br> $H_{23} =$ | $J_n e^{i(k_x - k_y)a}$ <br> $J_n e^{-i(k_x - k_y)a}$ | $J_n e^{-ik_y a}$ <br> $J_n e^{-ik_y a}$ | $J_n e^{ik_x a}$ <br> $J_n e^{-ik_x a}$ | $J_n$ <br> $J_n$ |

Here, $J_n$ is the NNN coupling strength. The bulk polarization can be calculate by using the Wilson-loop approach [2–5], thus it satisfies the relation

$$P_{x,y} = \frac{1}{2\pi} \sum_j^{N_{bg}} v_{x,y}^j (k_{y,x}) \quad (3)$$

Here, $v_x^j(k_y)$ is the $j$-th Wannier center along the x direction in reciprocal space, and $v_y^j(k_x)$ along the y direction. $N_{bg}$ is the total number of the bands below the gap of interest, in our case, there are $N_{bg} = 3$ bands [1] below the first band gap. The Wannier centers can be calculated by solving the eigenvalue problem of the Wannier Hamiltonian

$$H_{W_x}(k_y) = -\ln W_x(k_y), \quad H_{W_y}(k_x) = -\ln W_y(k_x)$$

$W_{x,y}$ are the Wilson-loop operators along the $x, y$ direction in the Brillouin zone, which are defined as

$$W_x(k_y) = F_x(k_x, k_y)F_x(k_x + \Delta k_x, k_y) \cdots F_x(k_x + 2\pi/a - \Delta k_x, k_y),$$
$$W_y(k_x) = F_y(k_x, k_y)F_y(k_x, k_y + \Delta k_y) \cdots F_y(k_x, k_y + 2\pi/a - \Delta k_y),$$

with $[F_x(k_x, k_y)]_{m,n} = \langle u_m(k_x, k_y) | u_n(k_x + \Delta k_x, k_y) \rangle$, $[F_y(k_x, k_y)]_{m,n} = \langle u_m(k_x, k_y) | u_n(k_x, k_y + \Delta k_y) \rangle$ and $\Delta k_x = \Delta k_y = 2\pi/aN$ (here, we choose $N = 500$, i.e. $501 \times 501$ points used for describing the first Brillouin zone). The Bloch functions $u_m(k_x, k_y)$ are calculated by using the tight-binding model Eq. (1) and satisfy $\langle u_m(k_x, k_y) | u_n(k_x, k_y) \rangle = \delta_{m,n}$. Therefore, the polarizations of different unit-cells can be numerically obtained and are summarized below:

|  | $v_x^1$ | $v_x^2$ | $v_x^3$ | $P_x$ | $v_y^1$ | $v_y^2$ | $v_y^3$ | $P_y$ |
|---|---|---|---|---|---|---|---|---|
| unit-cell I | 0 | $\pi$ | 0 | 1/2 | 0 | $\pi$ | 0 | 1/2 |
| unit-cell II | $\pi$ | 0 | $\pi$ | 0 | 0 | $\pi$ | 0 | 1/2 |
| unit-cell III | 0 | $\pi$ | 0 | 1/2 | $\pi$ | 0 | $\pi$ | 0 |
| unit-cell IV | $\pi$ | 0 | $\pi$ | 0 | $\pi$ | 0 | $\pi$ | 0 |

The quantized 2D Zak phase is related to the bulk polarization via the relation $\theta = 2\pi P$. The corresponding band gap is trivial when $\theta = 0$, while it is non-trivial when $\theta = \pi$. From the table above, one can see that unit-cell IV is trivial in both directions with $\theta = (0,0)$, while unit-cell I is non-trivial with $\theta = (\pi, \pi)$. Unit-cell III is non-trivial in the x direction and trivial in the y direction with $\theta = (\pi, 0)$, whereas unit-cell II is trivial in x and non-trivial in y with $\theta = (0, \pi)$.

In our experimental configuration as sketched in Fig. 1(b) of the main text, we thus describe our lattice as composed of four domains with unit-cells of types I-IV arranged as marked in panels (e,h).

**Supplementary section S2: Strong light-matter coupling in mCherry-filled cavities**

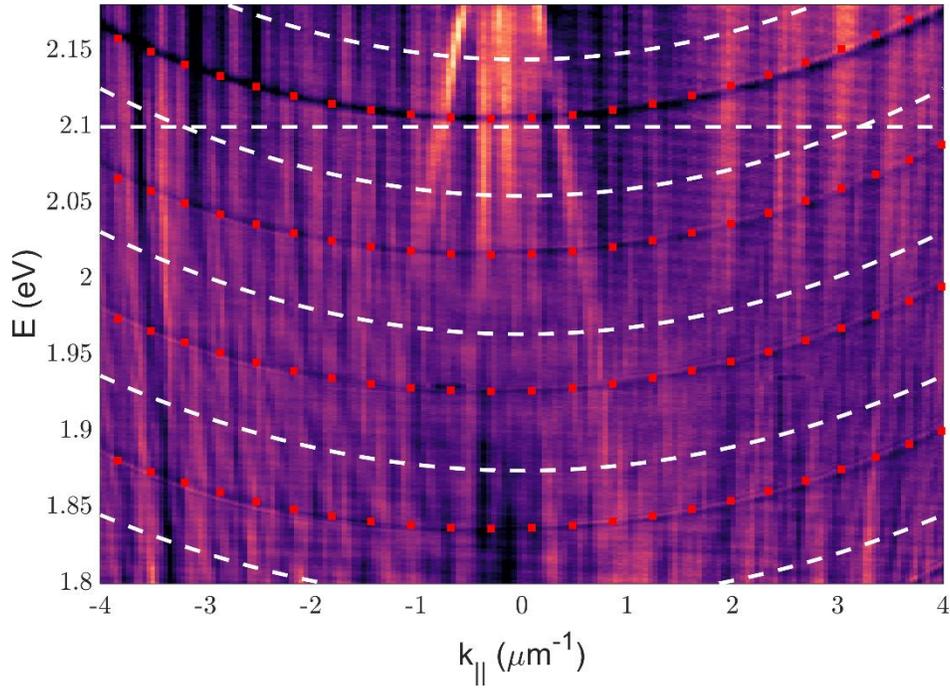

**Fig. S1: Strong light-matter coupling in mCherry-filled planar microcavity.** Polariton dispersion relation measured in white light reflection on a planar microcavity filled with mCherry. A coupled oscillator model (red squares) well accounts for the observed polariton modes with the uncoupled exciton at 2.1 eV in a cavity with $L_{cav}$=4.1 µm. The model yields a coupling strength of g=105 meV, i.e. a vacuum Rabi splitting of . $\hbar\omega_R$ = 210 meV. The dashed white lines mark the uncoupled photonic modes and the uncoupled exciton, respectively.

**Supplementary section S3: Tight-binding simulations including anisotropy in the hopping**

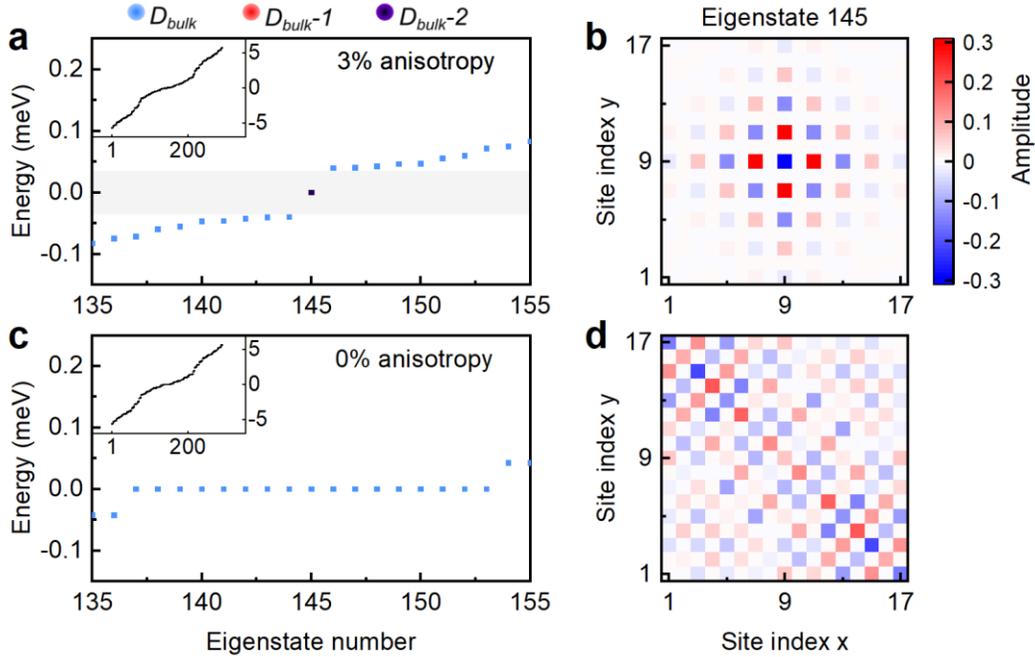

**Fig. S2: Eigenenergy spectrum of the 2D SSH lattice including hopping anisotropy but no next-nearest neighbor hopping. a** Eigenenergies of the 2D SSH lattice as shown in Fig. 1 (main text) with $\frac{v_x}{v_y} = \frac{w_x}{w_y} = 1.03$ and $2meV$ for the strong and $0.9 meV$ for the weak hopping elements along the $y$-direction, i.e. a 3% anisotropy in the hopping terms that is most likely caused by fabrication imperfections. An 80µeV wide gap opens around zero with one state pinned at its center. While this gap width is below our spectral imaging resolution in the experiment, we may still expect to be able to identify this state albeit with some admixture from close-by bulk states in the signal. Inset: In the absence of next-nearest neighbor hopping the full eigenspectrum is symmetric with respect to zero energy. The simulated states plotted in Fig. 3 of the main text correspond to eigenstates with numbers 145 (0D), 226 (1D) and 28 (2D bulk) in the spectrum shown in Fig. S2(a). **b** The state in the gap shows the expected sub-lattice polarization of the SSH state, is localized at the monomeric central defect and decays exponentially into the bulk. **c-d** For comparison, when hopping anisotropy is switched off, a large number of bulk states around zero energy become near-degenerate and no gap is present.

**Supplementary section S4: Cavity assembly**

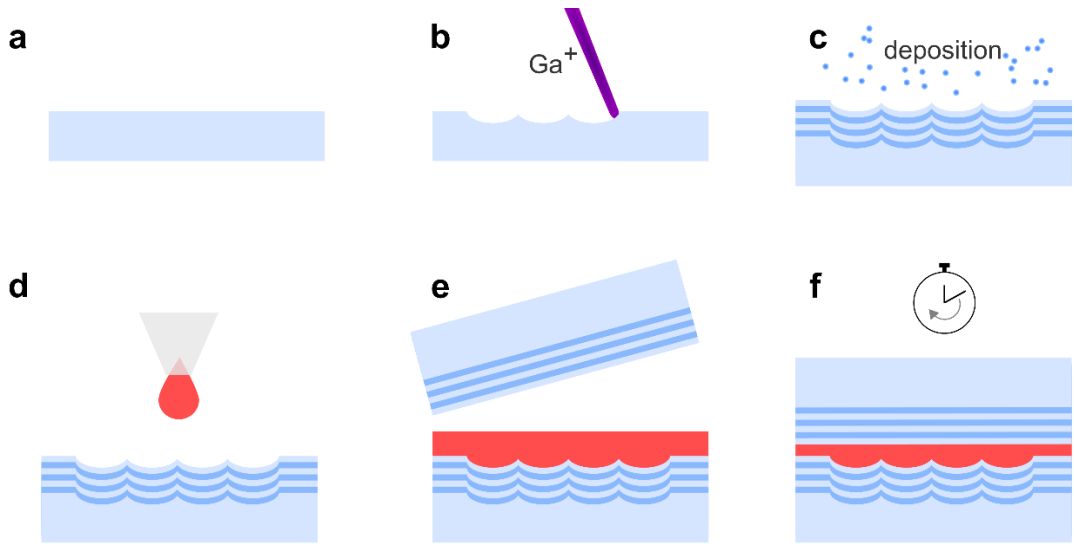

**Fig. S3: Schematic of cavity assembly. a** Planar substrate of eco – thin glass (SCHOTT). **b** Indentations in the shape of spherical caps with a depth of d=155nm are fabricated into the substrate using focused $Ga^+$ ion beam milling. **c** Alternating layers of $SiO_2$ and $TiO_2$ are evaporated onto the patterned (8 pairs) and planar substrate (10 pairs) to create dielectric mirrors. **d** A volume of 5 μL of a 175g/L fluorescent protein mCherry solution is drop-cast onto the patterned dielectric mirror. **e** A planar mirror is placed onto the mCherry-covered mirror. **f** The laminated cavity is left to dry for 48h in a temperature-stabilized environment at room-temperature, with a weight of 50g placed on the top of the mirrors.